\documentclass[aps, pra, 10pt, twocolumn, notitlepage,superscriptaddress,nofootinbib,longbibliography]{revtex4-2}
\usepackage{natbib}
\usepackage{graphicx}
\usepackage{dcolumn} 
\usepackage{bm}
\usepackage{mathtools}
\usepackage{amssymb}
\usepackage{enumitem}
\usepackage[svgnames,dvipsnames]{xcolor}
\usepackage[normalem]{ulem}
\usepackage[mathlines]{lineno}
\usepackage[titletoc, title]{appendix}
\usepackage{bbold}

\usepackage{hyperref}
    \hypersetup
    {   
        unicode=true,           
        pdftoolbar=true,        
        pdfmenubar=true,        
        pdffitwindow=false,     
        pdfstartview={FitH},    
        pdfnewwindow=true,      
        colorlinks=true,        
        linkcolor=Maroon,       
        citecolor=Cerulean,     
        filecolor=Maroon,       
        urlcolor=Cerulean       
    }

\newcommand{\bra}[1]{\langle #1 |}
\newcommand{\ket}[1]{| #1 \rangle}
\newcommand{\braket}[2]{\langle #1 | #2 \rangle}
\newcommand{\braop}[3]{\langle #1 | #2 | #3 \rangle}

\begin{document}

\title{Finite-temperature entanglement and coherence\\in asymmetric bosonic Josephson junctions}

\author{Cesare Vianello}
\email{cesare.vianello@phd.unipd.it}
\affiliation{Dipartimento di Fisica e Astronomia ``Galileo Galilei'', Università di Padova, Via Marzolo 8, I-35131 Padova, Italy}
\affiliation{Istituto Nazionale di Fisica Nucleare, Sezione di Padova, Via Marzolo 8, I-35131 Padova, Italy}
\author{Matteo Ferraretto}
\affiliation{Scuola Internazionale Superiore di Studi Avanzati, Via Bonomea 265, I-34136, Trieste, Italy}
\author{Luca Salasnich}
\affiliation{Dipartimento di Fisica e Astronomia ``Galileo Galilei'', Università di Padova, Via Marzolo 8, I-35131 Padova, Italy}
\affiliation{Istituto Nazionale di Fisica Nucleare, Sezione di Padova, Via Marzolo 8, I-35131 Padova, Italy}
\affiliation{Padua QTech Center, Università di Padova, Via Gradenigo 6/A, I-35131 Padova, Italy}
\affiliation{Istituto Nazionale di Ottica del Consiglio Nazionale delle Ricerche,
Via Carrara 1, I-50019 Sesto Fiorentino, Italy}

\begin{abstract} 
We investigate the finite-temperature properties of a bosonic Josephson junction composed of $N$ interacting atoms confined by a quasi-one-dimensional asymmetric double-well potential, modeled by the two-site Bose-Hubbard Hamiltonian. 
We compute numerically the spectral decomposition of the statistical ensemble of states, the thermodynamic and entanglement entropies, the population imbalance, the quantum Fisher information, and the coherence visibility. 
We analyze their dependence on the system parameters, showing in particular how finite temperature and on-site energy asymmetry affect the entanglement and coherence properties of the system. 
Moreover, starting from a quantum phase model which accurately describes the system over a wide range of interactions, we develop a reliable description of the strong tunneling regime, where thermal averages may be computed analytically using a modified Boltzmann weight involving an effective temperature. We discuss the possibility of applying this effective description to other models in suitable regimes.
\end{abstract}

\maketitle

\section{Introduction}

Quantum tunneling is at the basis of many physical phenomena, among which the Josephson effect stands out in atomic physics. 
One manifestation of this effect is the flow of a continuous current across a device constituted by two superconductors coupled by a weak link---a Josephson junction---without any voltage applied \cite{Josephson}. 
Although originally conceived in the context of superconductivity, a pair of weakly-coupled Bose-Einstein condensates confined by optical lattices that realize a quasi-one-dimensional double-well potential represents a bosonic Josephson junction (BJJ), providing a compelling platform for exploring fundamental quantum many-body physics \cite{Morsch, Albiez, Shin, Levy, Gati_Rev}. 
These systems exhibit rich dynamical behavior, such as Josephson oscillations, macroscopic quantum self-trapping, collapses and revivals, arising from the interplay of tunneling and interparticle interactions \cite{SmerziPRL, Smerzi, Williams, Tonel, Scazza_Josephson_necklace}. 
At zero temperature, both the ground state and the dynamics of BJJs have been extensively studied using a variety of numerical and mean-field approaches \cite{Milburn, Anglin, Trimborn, Diaz, Salasnich_Toigo, Dellanna, Kerkdyk, Pigneur, Wimberger}. However, real experimental systems are never truly at zero temperature. Thermal fluctuations and decoherence play a crucial role in determining the behavior of the junction, particularly in regimes where the thermal energy becomes comparable to the interaction or tunneling energies \cite{Pitaevski_Stringari, Gati, Gottlieb, JuliaPolls, Mazzarella_2012}. Understanding these effects is important not only for better interpreting experimental observations, but also for the advancement of technologies that rely on coherent bosonic systems, such as in the emerging field of `atomtronics' \cite{Amico_AVS, Amico_Rev}.

The paper aims to comprehensively characterize the finite-temperature properties of BJJs, with a special focus on entanglement and coherence. 
In Sec. \ref{sec:theory} we present the model Hamiltonian and Hilbert space. We then discuss how the system can be described by a quantum phase model over a wide range of interactions. From this we argue that in the strong tunneling regime, thermal averages of observables having a classical analogue can be computed analytically using a modified Boltzmann weight involving an effective temperature. 
In Sec. \ref{sec:res} we present exact numerical results for various properties of the system, obtained by diagonalizing the Hamiltonian. 
We analyze their dependence on temperature, interaction strength, energy asymmetry between the two wells, and number of particles, also extending previously reported results for the cases of zero-temperature and symmetric wells. We thus compare our effective analytical results with the exact ones, discussing the validity of the underlying approximations. 
Finally, in Sec. \ref{conclusion} we summarize the results and discuss possible extensions of this work, in particular the possibility of applying the effective semiclassical description in similar contexts.

\section{Theory}\label{sec:theory}

\subsection{The model Hamiltonian}

We consider a system of $N$ interacting bosons confined by a quasi-one-dimensional asymmetric double-well potential $V_\text{DW}(x)$, superimposed to a strong harmonic confinement in the transverse directions. 
The single-particle energy levels are arranged in quasi-degenerate doublets \cite{Muller}. 
Assuming that the potential barrier is high enough so that the energy gap between the first two doublets is much larger than the interaction, tunneling and thermal energies, the single-particle Hilbert space is restricted to the span of the first two eigenstates $\ket 0$ and $\ket 1$. 
Changing basis to the states $\ket L = (\ket 0 -\ket 1)/\sqrt{2}$ and $\ket R = (\ket 0 +\ket 1)/\sqrt{2}$, localized in the left and right well respectively, the system will be described by the two-site Bose-Hubbar model \cite{Leggett}
\begin{align}\label{HBH}
    \hat H &= \frac{U}{2}[\hat n_L(\hat n_L-1) + \hat n_R(\hat n_R-1)] + \frac{\varepsilon}{2} (\hat n_L - \hat n_R)\nonumber\\
    &-J(\hat a_L^\dag \hat a_R + \hat a_R^\dag \hat a_L).
\end{align}
Here $\hat a^\dag_{L(R)}$ and $\hat a_{L(R)}$ are bosonic creation and annihilation operators satisfying canonical commutation relations, and $\hat n_{L(R)} = \hat a_{L(R)}^\dag \hat a_{L(R)}$ are the number operators for each well; $U$ is the boson-boson interaction energy, with $U>0$ describing repulsive interaction and $U<0$ corresponding to attractive interaction; $\varepsilon$ is the on-site energy asymmetry; $J>0$ is the tunneling (hopping) energy between the two wells. The total number operator $\hat N = \hat n_L + \hat n_R$ commutes with $\hat H$ and is therefore a conserved quantity. The $N$-particle Hilbert space has dimension $N+1$ and is spanned by the basis of Fock states $\{\ket{n_L, n_R}\} = \{\ket{i, N-i}\}_{i=0,\dots,N}$. In this basis, the Hamiltonian is represented by a $(N+1)\times(N+1)$ real symmetric matrix, and its eigenstates are expanded as
\begin{equation}\label{eigen}
    \ket{E_n} = \sum_{i=0}^N c_i^{(n)}\ket{i, N-i},
\end{equation}
where the coefficients $c_i^{(n)}$ are real and $|E_n\rangle$ is normalized to unity. In the following we will study several properties of the model \eqref{HBH} in equilibrium at temperature $T \equiv (k_B\beta)^{-1}$, when the system is in the statistical ensemble of states determined by the density matrix
\begin{equation}\label{densmat}
    \hat\rho = \frac{e^{-\beta \hat H}}{Z} = \frac{1}{Z}\sum_{n=0}^N e^{-\beta E_n} \ket{E_n}\bra{E_n},
\end{equation}
with $Z$ being the canonical partition function. The thermal average of an operator $\hat Q$ is then $\langle\hat Q\rangle = \text{Tr}(\hat\rho \hat Q)$.

The model \eqref{HBH} is integrable via algebraic Bethe ansatz, see e.g. Ref. \cite{Links} and references therein. However, calculating matrix elements from the Bethe ansatz is very laborious. Since the dimension of the Hilbert space scales linearly with the number of particles, the direct numerical diagonalization of the Hamiltonian is computationally advantageous. The numerical results we obtain, although exact, do not offer a conceptual picture that may be applied beyond the specific model we are considering. We are therefore interested in developing an effective description that can provide valuable analytical insights. 

\subsection{Quantum phase model}

Taking the expectation value of the time-evolution equations $i\hbar \dot{\hat a}_{L(R)} = [\hat a_{L(R)}, \hat H]$ over the Glauber coherent state $\ket\alpha = \ket{\alpha_L}\otimes \ket{\alpha_R}$ defined by $\hat a_{L(R)} \ket{\alpha} = a_{L(R)} \ket{\alpha}$, and introducing the number-phase parametrization $a_{L(R)} = \sqrt{n_{L(R)}}e^{i\theta_{L(R)}}$ for the corresponding eigenvalues, one obtains two classical equations for the fractional population imbalance $z = (n_L-n_R)/N$ and the relative phase $\theta = \theta_R-\theta_L$ \cite{SmerziPRL, Smerzi},
\begin{subequations}\label{eq:Josephson_Smerzi}
\begin{align}
    \dot\theta &= \frac{2J}{\hbar}\frac{z}{\sqrt{1-z^2}}\cos\theta + \frac{UN}{\hbar}z + \frac{\varepsilon}{\hbar},\\
    \dot z &= -\frac{2J}{\hbar}\sqrt{1-z^2}\sin\theta.
\end{align}
\end{subequations}
Introducing the `canonical momentum' $\hbar k = \hbar Nz/2$, Eqs. (\ref{eq:Josephson_Smerzi}) can be regarded as the Hamilton equations $\hbar \dot\theta = \partial H/\partial k$, $\hbar \dot k = -\partial H/\partial \theta$ for the classical Hamiltonian
\begin{equation}\label{Hcl}
    H = U k^2 + \varepsilon k - JN\sqrt{1-\left(\frac{2k}{N}\right)^2}\cos\theta.
\end{equation}
We may expand the square root in Eq. \eqref{Hcl} to obtain $H = U k^2\bigl[1 + \frac{2J}{U N}(1 + \frac{k^2}{N^2} + \frac{2k^4}{N^4} + \cdots)\cos\theta] + \varepsilon k - JN\cos\theta$. Since both $k/N$ and $\cos\theta$ are bounded, assuming that $|U| \gg J/N$ the Hamiltonian simplifies to
\begin{equation}\label{HJcl}
    H_J = U k^2 + \varepsilon k - JN \cos\theta.
\end{equation}
The above condition will be satisfied for a wide range of $U$ and $J$ in the thermodynamic limit. 

Following Pitaevskii and Stringari \cite{Pitaevski_Stringari}, the classical $H_J$ can then be quantized by promoting the conjugate variables $\theta$ and $\hbar k$ to operators satisfying the commutation relation $[\hat \theta, \hbar \hat k] = i\hbar$. 
In the `$\theta$ representation', $\hat k = -i\partial/\partial\theta$ and
\begin{equation}\label{HJ}
    \hat H_J = -U\frac{\partial^2}{\partial\theta^2} - i\varepsilon\frac{\partial}{\partial\theta} - JN \cos\theta,
\end{equation}
acting on the space of $2\pi$-periodic wavefunctions. This quantum phase model (QPM) provides an effective description of the two-site Bose-Hubbard model \eqref{HBH}, and extends the model of Pitaevskii and Stringari to the case of asymmetric wells. Naively, we expect the QPM to be reliable in the interaction range suggested by its classical counterpart, namely $|U|\gg J/N$. We have verified that this is always true in the limit of a large number of particles, whereas for a system with few particles the naive condition of validity may not be sufficient to ensure accurate predictions, especially for those observables that depend on the relative phase $\theta$. More details on the QPM, how to compute thermal averages, and the accuracy of these results are reported in Appendix \ref{sec:PS}. Our main interest is to use it as the starting point for the semiclassical approximation presented below.

\subsection{Semiclassical approximation}\label{sec:semiclassical}

Consider the case of repulsive interaction, $U>0$. In the strong tunneling regime $U \ll JN$, the dynamical system modeled by Eq. \eqref{Hcl} undergoes small oscillations described by the linearized Eqs. \eqref{eq:Josephson_Smerzi} $\ddot\theta = -\omega^2 \theta$, $\ddot z = -\omega^2 z - 2J\varepsilon/\hbar^2$, where $\omega = \sqrt{2J(2J+UN)}/\hbar$ is the Josephson frequency \cite{SmerziPRL, Smerzi}. If the condition $U \gg J/N$ also holds, small oscillations follow the linearized equations of motion of the Hamiltonian $H_J$, whose characteristic frequency is
\begin{equation}\label{omegaj}
    \omega_J = \frac{\sqrt{2UJN}}{\hbar}.
\end{equation}
This is derived from the purely classical dynamics of $H_J$ and thus ignores quantum fluctuations of the phase. It has recently been shown, using the quantum effective action approach, that the oscillation frequency including first-order quantum corrections is \cite{Furutani}
\begin{equation}\label{Omegaj}
    \Omega_J = \omega_J\sqrt{1-\sqrt{\frac{U}{8JN}}}.
\end{equation}
At the level of the quantum Hamiltonian $\hat H_J$, this means that in the strong tunneling regime the statistics determining the distribution of states is essentially that of an harmonic oscillator with frequency $\Omega_J$. The probability distribution of positions and momenta of the quantum oscillator differs from the classical Boltzmann distribution only by the fact that the temperature $T$ is substituted by the effective temperature \cite{Landau, Melnikov}
\begin{equation}
    T_\text{eff} = \frac{\hbar\Omega_J}{2k_B}\coth\left(\frac{\hbar \Omega_J}{2k_B T}\right).
\end{equation}
In fact, the average energy of the quantum oscillator is $E_\text{ho} = \hbar \Omega_J/2 + \hbar \Omega_J/(e^{\beta\hbar\Omega_J}-1) = k_B T_\text{eff}$, and by the virial theorem the contribution of the kinetic and potential terms is the same, $\langle (\hbar\hat k)^2/2\rangle = \langle \Omega_J^2 \hat \theta^2/2\rangle = k_B T_\text{eff}/2$. 
In the limit $k_B T \gg \hbar \Omega_J$ we have $T_\text{eff}\to T$, and we recover the classical equipartition of energy. 

The above discussion provides us with a semiclassical approximation for the distribution of states in the regime $J/N \ll U \ll JN$, where we may calculate thermal averages as
\begin{equation}\label{semiclas}
    \langle \hat Q \rangle = \frac{1}{Z}\int_{-\frac{N}{2}}^{\frac{N}{2}}dk \int_{-\pi}^\pi d\theta\, Q(k,\theta) e^{-\beta_\text{eff}H_J(k,\theta)},
\end{equation}
with $Z = \int_{-N/2}^{N/2}dk \int_{-\pi}^\pi d\theta\,e^{-\beta_\text{eff}H_J(k,\theta)}$. It is clear that this possibility is conditioned on the existence of an observable $Q(k,\theta)$ that is the classical analogue of the operator $\hat Q$, which is not always the case. As we will discuss in the following sections, depending on the specific observable we are considering, the range of validity of Eq. \eqref{semiclas} may be subject to further conditions related to the number of particles $N$ and the temperature $T$.

\section{Thermal state of the BJJ}\label{sec:res}

In this section we characterize the thermal equilibrium state of the BJJ by inspecting the main observable quantities, such as the thermodynamic entropy, population imbalance, quantum Fisher information, and coherence visibility, as well as more theoretical quantities that provide a deeper understanding of the state, namely the entanglement entropy and the coefficients of the density matrix. We complement the numerical results obtained via the exact diagonalization of Eq. \eqref{HBH} with analytical results obtained from the semiclassical approach presented in Sec. \ref{sec:semiclassical}, discussing the validity of the underlying approximations. Each subsection is devoted to the discussion of a specific observable and contributes to the overall understanding of the finite-temperature properties of the system.

\subsection{Spectral decomposition}\label{sec:specdec}

Using Eq. \eqref{eigen}, we can write the density matrix \eqref{densmat} in the Fock basis as
\begin{equation}\label{rhodef}
    \hat \rho = \sum_{i,j=0}^N \rho_{ij}\ket{i, N-i}\bra{j, N-j},
\end{equation}
where
\begin{equation}
    \rho_{ij} = \frac{1}{Z}\sum_{n=0}^N e^{-\beta E_n} c_i^{(n)}c_j^{(n)}.
\end{equation}
In particular, the diagonal elements $\rho_{ii} \equiv \langle |c_i|^2 \rangle$ represent the average weights of the Fock states ${\ket{i, N-i}}$ in the statistical ensemble of states. At zero temperature, the diagonal elements reduce to $|c_i^{(0)}|^2$, that is the probability of observing the configuration $\ket{i, N-i}$ when the system is prepared in the ground state $\ket{E_0}$.

\begin{figure}[t]
    \centering
    \includegraphics[width=\linewidth]{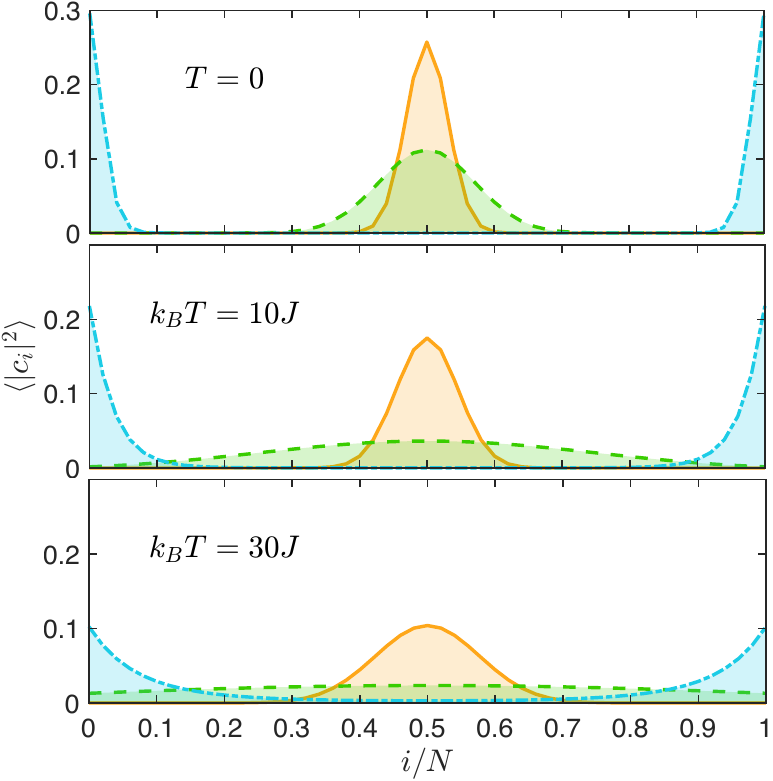}
    \caption{Spectral decomposition $\rho_{ii} = \langle |c_i|^2\rangle$ of the statistical ensemble of states for $\varepsilon=0$ as a function of $i/N$, plotted for $N=50$ and $U/J = 1.0$ (solid orange line), $0.0$ (dashed green line), $-0.2$ (dashed-dotted cyan line) at different temperatures.}
    \label{fig:1}
\end{figure}

The ground state of the symmetric model ($\varepsilon = 0$) is known to exhibit a rich phenomenology \cite{Greiner, Diaz, Salasnich_Toigo, Dellanna, Cirac, Dalvit, Mahmud, Jaask, Jaask2, Huang}, and is shown in the upper panel of Fig. \ref{fig:1}.
In absence of interaction, the ground state is the atomic coherent state ${\ket{\text{ACS}} = (1/\sqrt{N!})[(\hat a_L^\dag + \hat a_R^\dag)/\sqrt 2]^N\ket{0,0}}$, which corresponds to all $N$ particles occupying the single-particle ground state $\ket 0$ of the double-well potential, i.e. complete condensation. 
For repulsive interaction, by increasing $U/J$ there is a crossover from a superfluid-like regime ($U/J \ll N$), where the ground state is close to the atomic coherent state, to a Mott-like regime ($U/J \gg N$), where the ground state is incoherent and close to the separable twin Fock state $\ket{\text{FOCK}} = \ket{N/2, N/2}$. 
For attractive interaction, by decreasing $U/J$ we move from the superfluid-like regime to a `Schrödinger-cat regime', where the ground state tends to the symmetric entangled superposition of fully unbalanced Fock states $\ket{\text{CAT}_+} = (\ket{N,0} + \ket{0,N})/\sqrt 2$, also called `$N00N$ state', while the first excited state is close to the antisymmetric superposition $\ket{\text{CAT}_-} = (\ket{N,0} - \ket{0,N})/\sqrt 2$. 

In the attractive regime, there is a critical interaction $U_c/J = -2/N$ beyond which the energies of the ground state $E_0$ and the first excited state $E_1$ become quasi-degenerate (Fig. \ref{fig:3}a) \cite{Diaz, JuliaPolls, Cirac}. 
Moving deeper into the region $U<U_c$, the merging of energy levels occurs also within higher consecutive pairs, e.g. $E_3\simeq E_2$, $E_5\simeq E_4$, etc. 
In the thermodynamic limit $N\to\infty$, these quasi-degenerate doublets become degenerate. 
In particular, the ground state undergoes a quantum phase transition at $U=U_c$, signaled by a nonzero expectation value of the population imbalance in the two degenerate ground states, $\langle \hat z \rangle = \pm\sqrt{1-(U_c/U)^2}$. 

Let us now consider the finite-temperature mixed state. The qualitative behavior of the spectral decomposition $\langle |c_i|^2\rangle$ as a function of $U/J$ is similar to that of the ground state, however as one would expect, higher temperatures broaden the distribution of Fock states weights, leading to an increasingly less pure state (Fig. \ref{fig:1}). 
The transition between the superfluid-like and the cat-like regimes occurs around a characteristic value $U_\text{cat}/J<0$, which depends on the number of particles and on the temperature. For $N \gtrsim 10$, at fixed $k_BT/JN$ we have a simple power law, $U_\text{cat}/J = -\gamma N^{-\delta}$. For instance, the coefficients $\gamma$ and $\delta$ take the values $\gamma = 2.79$, $\delta = 1.07$ for $T = 0$; $\gamma = 2.00$, $\delta = 1.02$ for $k_BT/JN = 0.1$; $\gamma = 1.29$, $\delta = 0.98$ for $k_BT/JN = 0.5$.

\begin{figure}[t]
    \centering
    \includegraphics[width=\linewidth]{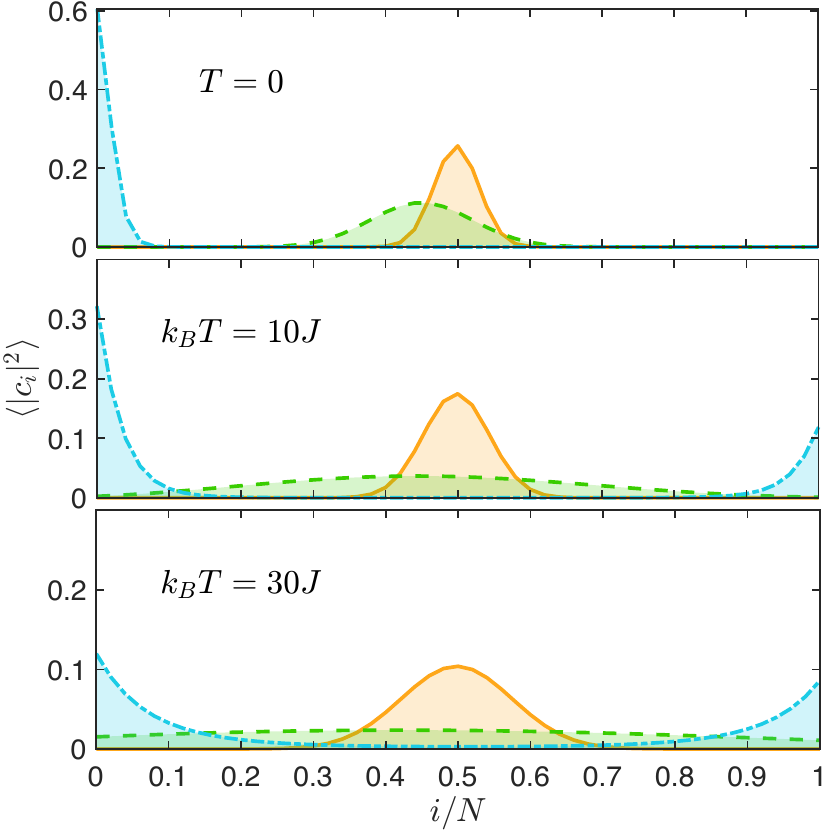}
    \caption{Spectral decomposition $\rho_{ii} = \langle |c_i|^2\rangle$ of the statistical ensemble of states for $\varepsilon/J = 0.2$ as a function of $i/N$, plotted for $N=50$ and $U/J = 1.0$ (solid orange line), $0.0$ (dashed green line), $-0.2$ (dashed-dotted cyan line) at different temperatures.}
    \label{fig:2}
\end{figure}

Introducing a nonzero $\varepsilon$ explicitly breaks the left-right symmetry and lifts the quasi-degeneracy of the two lowest eigenstates (Fig. \ref{fig:3}b). In this case no quantum phase transition occurs. Given $\varepsilon$, the left-right symmetry of $|c_i^{(0)}|^2$ is broken around a specific value of $U/J$, which is negative for weak asymmetry and positive for moderate and large asymmetry. For strong attraction the ground state tends to $\ket{0,N}$ for $\varepsilon > 0$ or $\ket{N,0}$ for $\varepsilon < 0$, while for strong repulsion the effect of the asymmetry is overcome, and the ground state tends in any case to $\ket{\text{FOCK}}$. At finite temperature, thermal fluctuations tend to give weight to all Fock states, attenuating the symmetry-breaking effects of $\varepsilon$ (Fig. \ref{fig:2}).

We note that at infinite temperature, the density matrix $\hat\rho = \mathbb 1/(N+1)$ describes a uniform mixture of the $N+1$ Fock states, and the quantum average of an operator $\hat Q$ is $\langle \hat Q\rangle_{\beta=0} = (N+1)^{-1}\sum_{i=0}^N \braop{i,N-i}{\hat Q}{i,N-i}$. Any diagonal operator in the Fock basis will thus average to the arithmetic mean of its eigenvalues, whereas any operator with off-diagonal character will average to zero. In the semiclassical approximation, the same average is given by $\langle \hat Q\rangle_{\beta=0}^{\text{cl}} = (2\pi N)^{-1}\int_{-N/2}^{N/2} dk \int_{-\pi}^\pi d\theta\,Q(k,\theta)$. While the two averages coincide in the large $N$ limit, at finite $N$ this is not guaranteed, but depends on the nature of the operator; in fact, they do not coincide when $\hat Q$ is probing genuine quantum correlations. This may place an upper bound on the range of temperatures over which the semiclassical approximation can give quantitatively accurate results at finite $N$. A concrete example will be discussed in Sec. \ref{sec:FI}.

\subsection{Thermodynamic entropy}

All equilibrium thermodynamic properties of the system can be derived from the thermodynamic entropy, that is the von Neumann entropy of the density matrix \eqref{rhodef} in units of the Boltzmann constant, 
\begin{align}\label{thent}
    S \equiv k_B S_\text{vN}(\hat\rho) &= -k_B \text{Tr}(\hat \rho \ln \hat\rho)\nonumber\\
    &= -k_B\sum_{n=0}^N\rho_n \ln \rho_n,
\end{align}
where $\rho_n = e^{-\beta E_n}/Z$. This is written in terms of the internal energy $E = \langle \hat H \rangle$ and the Helmholtz free energy $F = -k_B T \ln Z$ as $S = (E-F)/T$, that is the usual thermodynamic relation. 

\begin{figure}[t]
    \centering
    \includegraphics[width=\linewidth]{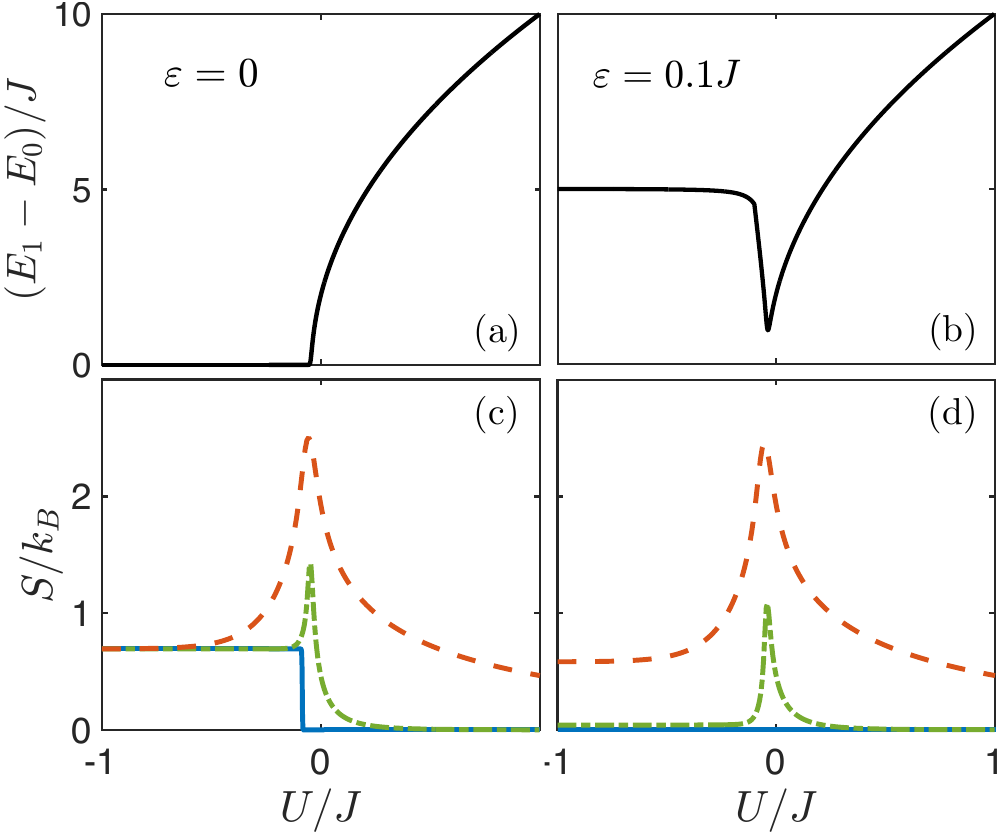}
    \caption{Energy gap between $\ket{E_0}$ and $\ket{E_1}$ (upper panels) and thermodynamic entropy (lower panels) as functions of $U/J$, plotted for $N=50$ and $k_BT/J = 10^{-10}$ (solid blue line), $1.0$ (dashed-dotted green line), $5.0$ (dashed orange line). The energy asymmetry is $\varepsilon = 0$ (left panels) and $\varepsilon/J = 0.1$ (right panels).}
    \label{fig:3}
\end{figure}

At zero temperature and finite $N$, the entropy is $S = 0$ for any $U$, $J$ and $\varepsilon$, as the ground state is non-degenerate. For $\varepsilon = 0$, in the thermodynamic limit, where the ground state is two-fold degenerate for $U<U_c$, the zero-temperature entropy is $S = \Theta(U_c-U) k_B \ln 2$, where $\Theta$ is the unit step function.
The same dependence on $U$ is observed at finite $N$ for infinitesimal temperatures; $S$ jumps from zero to $k_B \ln 2$ in correspondence of the value of $U/J$ such that $E_1-E_0 \simeq k_BT$, as thermal fluctuations make both the quasi-degenerate states $\ket{E_0}$ and $\ket{E_1}$ accessible (Fig. \ref{fig:3}c). 
At finite temperature, $S$ is clearly larger than at zero temperature; it has a global maximum for a value of $U/J$, we call it $U_\text{th}/J$, close to $U_c/J$, and it is asymptotic to $k_B \ln 2$ for $U/J \to -\infty$ and to zero for $U/J \to +\infty$. This is a consequence of the fact that the gap $E_2-E_1$ grows almost linearly with $|U|/J$ in the attractive regime, so that in the limit $U/J \to -\infty$ thermal fluctuations can make only the first two energy levels accessible, while in the repulsive regime $E_1-E_0$ grows monotonically with $U/J$, and in the limit $U/J \to +\infty$ only the ground state is accessible. Introducing the energy asymmetry $\varepsilon \neq 0$, whose sign is irrelevant for the entropy, quasi-degeneracies are lifted, and the asymptotic value of $S$ for $U/J\to-\infty$ acquires a dependence on $T$ as well as on $|\varepsilon|$ (Fig. \ref{fig:3}d).

\begin{figure}
    \centering
    \includegraphics[width=\linewidth]{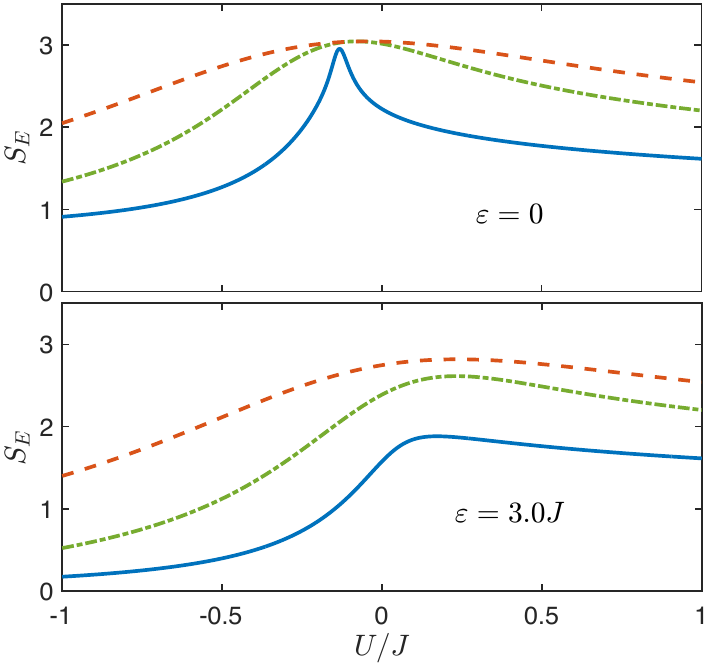}
    \caption{Entanglement entropy as a function of $U/J$, plotted for $N=20$ and $k_BT/J = 0$ (solid blue line), $10.0$ (dashed-dotted green line), $20.0$ (dashed orange line). The energy asymmetry is $\varepsilon = 0$ (upper panel) and $\varepsilon/J = 3.0$ (lower panel).}
    \label{fig:4}
\end{figure}

\subsection{Entanglement entropy}

The entanglement between the two wells can be characterized in terms of the reduced density matrices $\hat\rho_{L(R)} = \text{Tr}_{R(L)} \hat\rho$, obtained as the partial traces of the full density matrix \eqref{rhodef} over the degrees of freedom of each well. We note that $\hat \rho_L = \hat \rho_R$, which is given explicitly by $\hat\rho_R = \sum_{i,i',j=0}^N \rho_{ij}(\bra{i'}_L\otimes\text{id}_R)\ket{i, N-i}\bra{j,N-j}(\ket{i'}_L\otimes \text{id}_R) = \sum_{i,i',j=0}^N \rho_{ij} \delta_{ii'}\delta_{ji'}\ket{i, N-i}\bra{j, N-j}$, namely
\begin{equation}
    \hat\rho_R =\sum_{i=0}^N \rho_{ii} \ket{i, N-i}\bra{i,N-i} = \sum_{n=0}^N \rho_n \hat \rho^{(n)}_\text{diag},
\end{equation}
where $\hat\rho^{(n)}_\text{diag} = \sum_{i=0}^N|c_i^{(n)}|^2\ket{i, N-i}\bra{i,N-i}$. Thus the subsystems $L$ and $R$ have the same reduced density matrix, which is diagonal; thermal equilibrium is reflected in the fact that $\hat\rho_R$ is a statistical ensemble of the diagonal density matrices $\hat \rho^{(n)}_\text{diag}$ with Boltzmann weights $\rho_n$. We have entanglement when the entanglement entropy $S_E \equiv S_\text{vN}(\hat \rho_R)$, i.e. the von Neumann entropy of the reduced density matrix \cite{Amico}, is nonzero. This is given by
\begin{align}\label{entent}
    S_E = -\sum_{i=0}^N \langle|c_i|^2\rangle\ln\langle|c_i|^2\rangle.
\end{align}
At zero temperature, where $\hat\rho_R = \hat \rho^{(0)}_\text{diag}$, it reduces to 
\begin{equation}
    S_E = -\sum_{i=0}^N |c_i^{(0)}|^2 \ln |c_i^{(0)}|^2\qquad (T=0).
\end{equation} 
For a system with $N$ particles, $S_E$ takes values in the interval $[0, \ln(N+1)]$. It is zero if and only if $\hat \rho_R$ describes a pure state, while it is $\ln(N+1)$ when the system is in the maximally entangled state described by $\hat\rho_R = \mathbb 1/(N+1)$.

\begin{figure}[t]
    \centering
    \includegraphics[width=\linewidth]{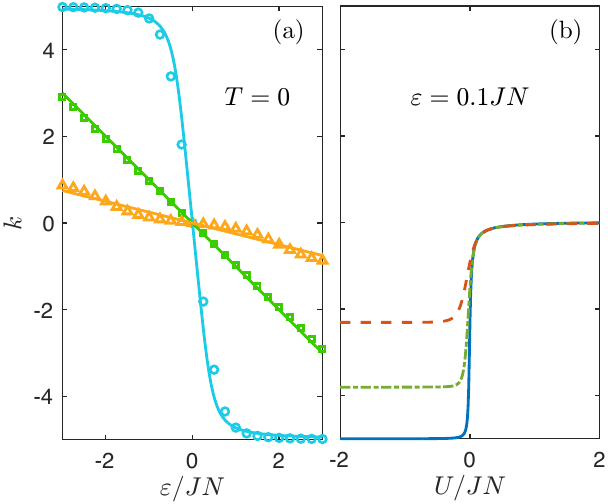}
    \caption{(a) Population imbalance at $T=0$ as a function of $\varepsilon/JN$, plotted for $N=10$ and $U/JN = 0.05$ (cyan circles), $0.5$ (green squares), $2.0$ (orange triangles). The continuous lines are the corresponding semiclassical results [Eq. \eqref{ksemi}]. (b) Population imbalance at $\varepsilon/JN = 0.1$ as a function of $U/JN$, plotted for $N=10$ and $k_BT/JN = 0$ (solid blue line), $0.5$ (dashed-dotted green line), $1.0$ (dashed orange line).}
    \label{fig:5}
\end{figure}

The relationship between entanglement entropy and thermodynamic entropy is a matter of great interest, and in generic (i.e. non-integrable) isolated systems is closely related to the eigenstate thermalization hypothesis \cite{Santos, Deutsch, Bhakuni_2020}. 
In our case, some insights can be obtained from the properties of the von Neumann entropy \cite{Nielsen}, whose subadditivity and concavity imply that
\begin{equation}
    \frac{S}{2k_B} \le S_E \le \frac{S}{k_B} + \sum_{n=0}^N \rho_n S_\text{vN}(\hat\rho^{(n)}_\text{diag}).
\end{equation}
Factoring out the partition function from Eq. \eqref{entent} and expressing it as $Z = e^{-\beta(E-TS)}$, we can also write
\begin{equation}\label{Sent-S}
    S_E = \frac{S}{k_B} -\beta E + e^{\beta(E-TS)}S_\text{vN}\Bigl(\textstyle \sum\limits_{n=0}^N e^{-\beta E_n}\hat\rho^{(n)}_\text{diag}\Bigr).
\end{equation}
Although this identity does not provide new information with respect to Eqs. \eqref{thent} and \eqref{entent}, it explicitly shows that in the large temperature limit $S_E \simeq S/k_B$. Furthermore, $S_E$ and $S/k_B$ take the same asymptotic values for $U/J \to \pm\infty$.

The entanglement entropy is plotted in Fig. \ref{fig:4} for several values of the temperature and two different values of $\varepsilon$. It has a maximum for a specific value of $U/J$,  which we denote as $U_\text{ent}/J$, where it is very close to the maximum achievable value $\ln(N+1)$, which for $N=20$ is about 3.04. This characteristic value of the interaction corresponds to a `maximally widespread' distribution of Fock state weights. At $T=\varepsilon=0$, $U_\text{ent}/J$ is very close to $U_\text{th}/J$ and $U_c/J$ (Figs. \ref{fig:3} and \ref{fig:4}). A nonzero $\varepsilon$ has the effect of increasing all these values. Higher temperature, besides leading to a larger entanglement entropy at any value of the interaction, also leads to larger values of $U_\text{ent}/J$ and $U_\text{th}/J$, which thus are shifted with respect to $U_c/J$, and to larger differences $|U_\text{ent}- U_\text{th}|/J$. By increasing the number of particles, the magnitude of these effects is reduced, and the characteristic values $U_c/J$, $U_\text{th}/J$ and $U_\text{ent}/J$ all tend to zero. In the thermodynamic limit, therefore, they match exactly and are equal to zero. A qualitative understanding of these shifts comes from thinking about the effect of $T$ and $\varepsilon$ on the distribution of Fock states weights (Sec. \ref{sec:specdec}). Starting from $U^{(0)}_\text{ent}/J<0$ at $T=\varepsilon=0$, both an increase in $T$ and an increase in $|\varepsilon|$ tend to reshape the distribution, making it less spread out. A slight increase in $U/J$ counterbalances this change, so that the maximum entanglement entropy will occur at a value $U_\text{ent} > U_\text{ent}^{(0)}$. Clearly, further increasing $U/J$ will start to narrow the distribution, reducing again the entanglement entropy, which explains the non-monotonic behavior observed in Fig. \ref{fig:4}. At large temperatures, the distribution of Fock states weights will in any case be very spread out, so that changing the interaction strength will have minimal effects; this explains the progressive flattening of the curves in the Fig. \ref{fig:4} as $T$ increases.

\subsection{Population imbalance}\label{popim}

The population imbalance can be characterized by the expectation value of the relative number operator
\begin{equation}\label{kexact}
    k = \langle \hat k \rangle = \frac{\langle \hat n_L - \hat n_R \rangle}{2} = \sum_{j=0}^N (j-N/2) \langle |c_j|^2 \rangle,
\end{equation}
which is bounded in the interval $[-N/2,\,N/2]$. 
In the QPM, $\hat k = -i \partial /\partial\theta$ and its average can be computed as discussed in Appendix \ref{sec:PS}. 
Both these definitions require a numerical calculation. However, in the strong tunneling regime we can obtain an analytic result using the semiclassical approach discussed in Sec. \ref{sec:semiclassical}; from Eq. \eqref{semiclas} we get
\begin{widetext}
\begin{equation}\label{ksemi}
    k = -\frac{\varepsilon}{2U} + \frac{e^{\beta_\text{eff}N\varepsilon}-1}{\sqrt{\pi U\beta_\text{eff}}}\frac{e^{-\beta_\text{eff}N^2U(\varepsilon/NU+1)^2/4}}{\text{erf}[\sqrt{\beta_\text{eff}N^2 U/4}(\varepsilon/NU+1)]-\text{erf}[\sqrt{\beta_\text{eff}N^2 U/4}(\varepsilon/NU-1)]},
\end{equation}
\end{widetext}
where erf$(x)$ denotes the error function. 
The slope of $k(\varepsilon)$ at $\varepsilon = 0$ represents the susceptibility of the particle imbalance to the on-site energy asymmetry $\varepsilon$, and can be regarded as a sort of capacitance of the bosonic Josephson junction in the linear-response regime. 
Deriving Eq. \eqref{ksemi} with respect to $\varepsilon$, we get
\begin{equation}
    k'(0) = - \frac{1}{2U} + \frac{N\beta_\text{eff}}{2\sqrt{\pi U \beta_\text{eff}}}\frac{e^{-\beta_\text{eff}N^2U/4}}{\text{erf}(\sqrt{\beta_\text{eff}N^2U/4})}.
\end{equation}

The semiclassical solution \eqref{ksemi} as a function of $\varepsilon/JN$ is compared with the exact result in Fig. \ref{fig:5}a. 
First notice that due the left-right symmetry, $k$ is strictly zero for $\varepsilon = 0$ and finite $N$.
Although we are making the comparison between semiclassical and exact results at $T=0$, where the system is maximally far from its classical limit and quantum effects are prevalent, our semiclassical approach works very well in the expected regime $1/N^2 \ll U/JN \ll 1$. 
Remarkably, even for $U/JN=2.0$, although the semiclassical solution is not able to trace the small-amplitude oscillations observed in the exact result, it correctly approximates the behavior of $k$.

In Fig. \ref{fig:5}b the population imbalance is plotted as a function of the boson-boson interaction. In the repulsive case, $k$ saturates to $\text{sgn}(-\varepsilon)N/2$ when the energy asymmetry dominates over the boson-boson interaction. The same holds at finite temperature when the energy asymmetry dominates also over the thermal energy. Conversely, $k$ goes to zero when the interaction dominates over the energy asymmetry and the temperature. 
The attractive regime is completely different from the repulsive one, for a weak attraction among the bosons is sufficient to saturate the population imbalance to a value whose modulus increases with $|\varepsilon|$ and decreases with $T$, reaching $N/2$ at $T=0$.

\subsection{Quantum Fisher information}\label{sec:FI}

Another signature of entanglement is the (quantum) Fisher information for the relative phase between the condensates in the two wells \cite{Braunstein, Pezze, Gert, Salasnich_Toigo, Dellanna, Mazzarella_2012}. 
Considering the relative number operator $\hat k$ as the generator of phase shifts, the Fisher information is defined as
\begin{align}
    \mathcal{I} &= \frac{4}{N^2} \left( \langle\hat k^2\rangle-\langle\hat k\rangle^2 \right) = \frac{4 \Delta k^2}{N^2},
\end{align}
and takes values in the interval $[0,1]$. For pure states, e.g. the ground state when the system is at zero temperature, a sufficient (but not necessary) condition for particle entanglement is $\mathcal I > 1/N$ \cite{Pezze}. In addition to being a flag for entanglement, $\mathcal I$ carries information about quantum fluctuations in the relative number of particles and relative phase. 
Since these are conjugate variables satisfying $[\hat \theta, \hat k] = i$, we can use the Fisher information to estimate a lower bound for the quantum fluctuation of the phase $\Delta \theta^2 \equiv \langle \theta^2 \rangle - \langle \theta \rangle^2$ via the uncertainty principle $\Delta \theta \Delta k \geq 1/2$, which yields $\Delta \theta^2 \geq 1/(N^2 \mathcal I)$. Consequently, $\Delta \theta^2$ becomes extremely large as $\mathcal I\to 0$, while it may be as small as $1/N^2$ for $\mathcal I \to 1$.

The average of $\hat k^2$ is $\langle \hat k^2\rangle = \sum_{j=0}^N (j-N/2)^2\langle|c_j|^2\rangle$, which combined with Eq. \eqref{kexact} gives the exact result for $\mathcal I$. In the QPM $\hat k^2 = -\partial^2/\partial\theta^2$, and its average can be computed as discussed in Appendix \ref{sec:PS}. Within the semiclassical approach [Eq. \eqref{semiclas}] we obtain the analytical result
\begin{widetext}
\begin{equation}\label{k2av}
    \langle \hat k^2\rangle = \frac{\varepsilon^2}{4U^2} + \frac{1}{2U\beta_\text{eff}}+\frac{N}{2\sqrt{\pi U \beta_\text{eff}}}\frac{[1+e^{\beta_\text{eff}N\varepsilon}+ (\varepsilon/NU)(e^{\beta_\text{eff}N\varepsilon}-1)]e^{-\beta_\text{eff}N^2U(\varepsilon/NU+1)^2/4}}{\text{erf}[\sqrt{\beta_\text{eff}N^2 U/4}(\varepsilon/NU-1)]-\text{erf}[\sqrt{\beta_\text{eff}N^2 U/4}(\varepsilon/NU+1)]}.
\end{equation}
\end{widetext}
Combining this result with Eq. \eqref{ksemi}, we readily get a semiclassical expression for $\mathcal I$. 
Notice that for $\varepsilon = 0$, Eq. \eqref{k2av} simplifies to
\begin{equation}\label{Isemi}
    \langle \hat k^2(0)\rangle = \frac{1}{2U\beta_\text{eff}}-\frac{N}{2\sqrt{\pi U \beta_\text{eff}}}\frac{e^{-\beta_\text{eff}N^2U/4}}{\text{erf}(\sqrt{\beta_\text{eff}N^2U/4})}.
\end{equation}
This result is consistent with the fluctuation-dissipation relation
\begin{equation}
    \langle\hat k^2(0)\rangle = -k_B T_\text{eff}k'(0),
\end{equation}
which states that the Fisher information (i.e. the fluctuation of the particle imbalance) at $\varepsilon = 0$ is proportional to the junction capacitance (i.e. the response of the particle imbalance to an externally applied on-site potential) through the effective temperature.

\begin{figure}[t!]
    \centering
    \includegraphics[width=\linewidth]{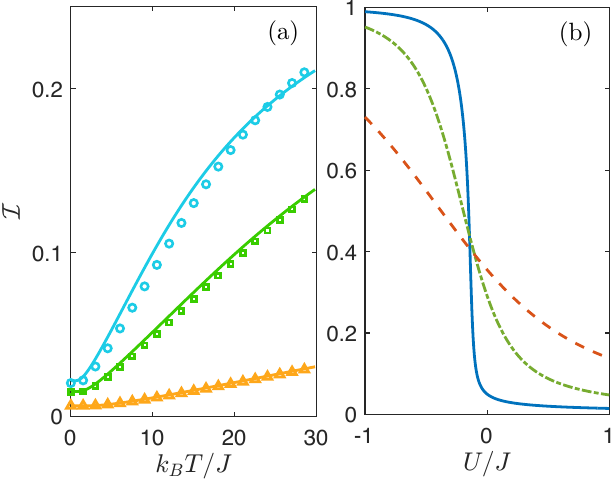}
    \caption{(a) Quantum Fisher information at $\varepsilon=0$ as a function of $k_BT/J$, plotted for $N=20$ and $U/J = 0.5$ (cyan circles), $1.0$ (green squares), $5.0$ (orange triangles). The continuous lines are the corresponding semiclassical results [Eq. \eqref{semiclas}]. (b) Quantum Fisher information at $\varepsilon = 0$ as a function of $U/J$, plotted for $N=20$ and $k_BT/J = 0$ (solid blue line), $10.0$ (dashed-dotted green line), $30.0$ (dashed orange line).}
    \label{fig:6}
\end{figure}

The semiclassical solution \eqref{Isemi} as a function of $k_BT/J$ is compared with the exact result in Fig. \ref{fig:6}a. Once again we observe good agreement in the expected range of interaction. The solid cyan line shows some deviations as the temperature is increased; in particular, we see that the semiclassical result is going to underestimate the exact result in the limit $T\to\infty$. Being a correlator, $\langle \hat k^2 \rangle$ is in fact one of those averages for which the semiclassical description becomes inaccurate above a certain temperature (see the discussion of Sec. \ref{sec:specdec}). The semiclassical result at infinite temperature is $\langle \hat k^2\rangle_{\beta=0}^\text{cl} = N^2/12$, whereas the exact result is $\langle \hat k^2\rangle_{\beta=0} = N^2(1+2/N)/12$. The relative error at infinite temperature is therefore $2/(N+2)$, that is about $9\%$ for $N=20$. As expected, the semiclassical approximation will become increasingly accurate as $N$ increases.

In Fig. \ref{fig:6}b the Fisher information is plotted as a function of the boson-boson interaction for $\varepsilon = 0$. We notice that at zero temperature, $\mathcal I$ is exactly equal to $1/N$ for vanishing boson-boson interaction, while for $U/J$ large and positive it is close to zero. 
This can be understood from the fact that in such regime the ground state is close to the separable state $\ket{\text{FOCK}}$, so that no large fluctuations of the number of particles in each well are expected. 
At finite temperature, thermal fluctuations provide a larger probability to other Fock states with a slightly imbalanced population, thus increasing $\mathcal{I}$ (see also Fig. \ref{fig:6}a).
Conversely, for $U/J$ large and negative $\mathcal I$ is close to unity. In this regime the ground state is close to the entangled state $\ket{\text{CAT}_+}$, so that in a series of repeated measurements, the observed relative number of particles in the two wells would oscillate violently between zero and $N$. At finite temperature, states with a less prominent particle imbalance acquire a larger probability, resulting in a reduced value of $\mathcal{I}$. In practice, temperature has the tendency of smoothing out the fluctuations of the relative number of particles, so that $\mathcal I$ varies more slowly as a function of $U/J$ (Fig. \ref{fig:6}b).

Introducing a nonzero $\varepsilon$ has an important effect for $U<0$. Since $\langle\hat k\rangle$ quickly saturates to a nonzero value (see Fig. \ref{fig:5}b), $\mathcal I$ is asymptotic to a value smaller than unity for strong attraction. At fixed $|\varepsilon|$, higher temperature leads to a larger asymptotic value, whereas keeping the temperature fixed, by increasing $|\varepsilon|$ the asymptotic value becomes smaller.

\subsection{Coherence visibility}

The coherence of our system can be characterized in terms of the momentum distribution $n(p) = N^{-1}\langle \hat\Psi^\dag(p)\hat \Psi(p)\rangle$, where $\hat \Psi(p)$ is the Fourier transform of the field operator $\hat \Psi(x) = \hat a_L \psi_L(x) + \hat a_R \psi_R(x)$, with $\psi_{L(R)}(x) = \braket{x}{L(R)}$. 
For symmetric or weakly asymmetric wells, we take $\psi_{L(R)}(x) = \psi(x\pm d/2)$, where $d$ is the distance between the two minima of $V_\text{DW}(x)$ \cite{Stringari_coherence, Ferrini}, which are the Wannier functions of the double well. We thus obtain $\hat \Psi(p) = \psi(p)(\hat a_L e^{ipd/2\hbar} + \hat a_R e^{-ipd/2\hbar})$, and therefore
\begin{equation}
    n(p) = n_0(p) \biggl(1+\frac{\langle \hat a^\dag_L \hat a_R\rangle}{N}e^{-ipd/\hbar} + \frac{\langle \hat a^\dag_R \hat a_L\rangle}{N}e^{ipd/\hbar}\biggr),
\end{equation}
where $n_0(p)=|\psi(p)|^2$ is the momentum distribution of each condensate. Since the eigenstates of $\hat H$ are real in the Fock basis, $\langle \hat a^\dag_L \hat a_R \rangle = \langle \hat a_R^\dag \hat a_L\rangle$ is also real. The momentum distribution is then
\begin{equation}\label{momdis}
    n(p) = n_0(p) [1 + \alpha \cos(pd/\hbar)]
\end{equation}
and exhibits interference fringes of period $\Delta p = 2\pi\hbar/d$. 
The parameter
\begin{align}
    \alpha &= \frac{2\langle \hat a_L^\dag \hat a_R\rangle}{N}\nonumber\\
    &= \frac{1}{Z}\sum_{n,j=0}^N e^{-\beta E_n}\frac{2}{N}\sqrt{(1+j)(N-j)}c_{j+1}^{(n)}c_j^{(n)}
\end{align}
represents the amplitude of the interference fringes and is called coherence visibility. This is directly related to the occupation of the single-particle ground state $\ket 0$. In fact ${\hat a_0 = (\hat a_L + \hat a_R)/\sqrt 2}$ and the associated occupation number is $\hat n_0 = (\hat N + \hat a_L^\dag \hat a_R + \hat a_R^\dag \hat a_L)/2$; the condensate fraction is therefore
\begin{equation}
    \frac{\langle \hat n_0\rangle}{N} = \frac{1+\alpha}{2}.
\end{equation}

\begin{figure}
    \centering
    \includegraphics[width=\linewidth]{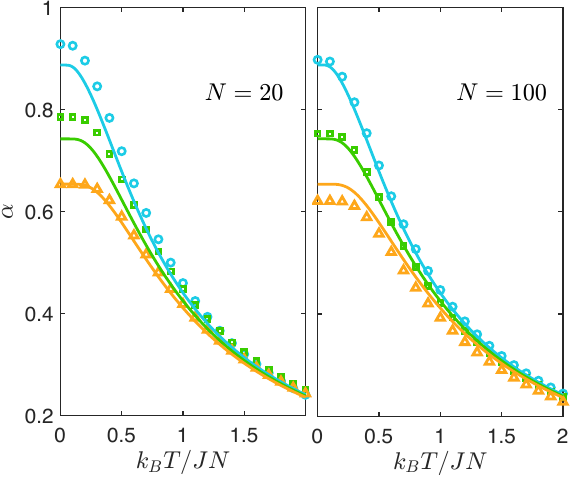}
    \caption{Coherence visibility at $\varepsilon=0$ as a function of $k_BT/JN$, plotted for $N=20$ (left panel) and $N=100$ (right panel), and $U/JN = 0.1$ (cyan circles), $0.5$ (green squares), $1.0$ (orange triangles). The continuous lines are the corresponding semiclassical results [Eq. \eqref{alphasemiclas}].}
    \label{fig:7}
\end{figure}

In the framework of the QPM, where the operator $\hat a_L^\dag \hat a_R + \hat a_R^\dag \hat a_L$ is averaged on the coherent state $\ket\alpha$, the coherence visibility is the expectation value of the cosine of the relative phase,
\begin{equation}
    \alpha = \langle \cos{\hat\theta} \rangle,
\end{equation}
and can be computed as discussed in Appendix \ref{sec:PS}.
Within the semiclassical approach [Eq. \eqref{semiclas}] we then obtain the analytical result
\begin{equation}\label{alphasemiclas}
    \alpha = \frac{I_1(JN\beta_\text{eff})}{I_0(JN\beta_\text{eff})},
\end{equation}
where $I_n(x)$ are modified Bessel functions of the first kind. Here the dependence on $U$ is contained only in $\beta_\text{eff}$, meaning that in the purely classical limit $\beta_\text{eff}\to\beta$ the coherence visibility does not depend on the boson-boson interaction but only on the tunneling energy and the number of particles.
In the high-temperature limit $JN\beta_{\text{eff}} \simeq JN\beta \ll 1$, expanding the Bessel functions we obtain $\alpha \simeq (2k_BT/JN)^{-1}$, which describes the hyperbolic tails observed in Fig. \ref{fig:7}. As expected, at infinite temperature the coherence visibility is zero, corresponding to $\langle \hat n_0\rangle/N = 1/2$, i.e. equal occupation of the two single-particle levels $\ket 0$ and $\ket 1$.

\begin{figure}
    \centering
    \includegraphics[width=\linewidth]{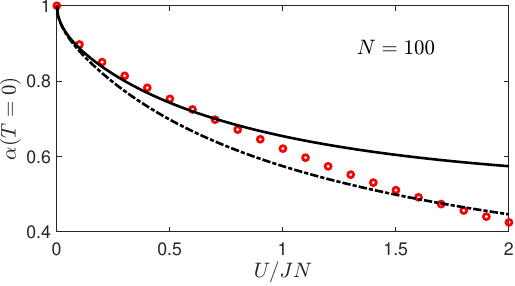}
    \caption{Coherence visibility at $\varepsilon=0$ and $T=0$ as a function of $U/JN$, plotted for $N=100$. The exact result (red circles) is compared with the semiclassical approximation based on the oscillation frequency $\omega_J = \sqrt{2UJN}/\hbar$ (dashed-dotted line) and with the improved semiclassical approximation based on the oscillation frequency including first order quantum corrections, $\Omega_J = \omega_J\sqrt{1-\sqrt{U/8JN}}$ (solid line).}
    \label{fig:9}
\end{figure}

The semiclassical solution \eqref{alphasemiclas} as a function of $k_BT/JN$ is compared with the exact result in Fig. \ref{fig:7}, for two different values of $N$. For $N=100$, we observe good agreement at all temperatures when the interaction is in the expected range (the solid orange line corresponds to $U/JN = 1.0$, i.e. the upper limit of the validity range, and in fact begins to show clear deviations from the exact result). For a smaller number of particles, $N=20$, the agreement remains decent at higher temperatures, while it is poor at low temperatures, even if the interaction is in the expected validity range. 
The reason for this difficulty, which has not been encountered for previous observables, is twofold. First, for repulsive interaction quantum fluctuations in the relative number of particles are typically small, while quantum fluctuations in the relative phase are very large. Since the coherence visibility is a function of the relative phase, while the other observables we considered were functions of the relative number, taking into account quantum fluctuations is decidedly more important for the visibility. 
This has already been partly addressed by including first-order quantum corrections in the frequency $\Omega_J$ [Eq. \eqref{Omegaj}], and we see that it is sufficient to guarantee good agreement with the exact result in the case $N=100$. In fact, if we compare the exact zero-temperature visibility with the ones obtained from the semiclassical approach using the purely classical frequency $\omega_J$ and the quantum-corrected frequency $\Omega_J$ (Fig. \ref{fig:9}), we see that the second one significantly improves the agreement with the exact result for $U/JN \ll 1$ (the improvement would not be as notable for the population imbalance and the Fisher information). Reducing the number of particles, this is no longer sufficient to obtain quantitative agreement with the exact result, due to the limitations of the underlying QPM, discussed in Appendix \ref{sec:PS}.

\begin{figure}[t]
    \centering
    \includegraphics[width=\linewidth]{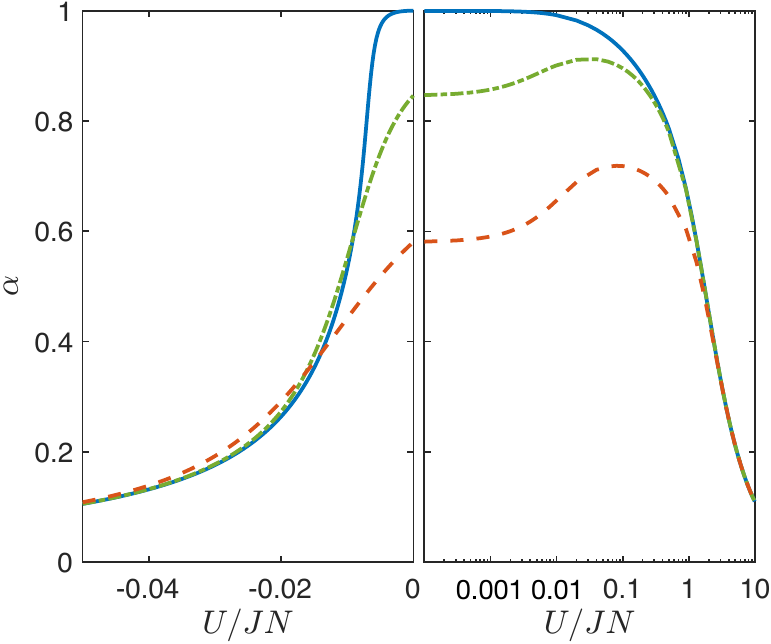}
    \caption{Coherence visibility at $\varepsilon=0$ as a function of $U/JN$, plotted for $N=20$ and $k_BT/JN = 0$ (solid blue line), $0.2$ (dashed-dotted green line), $0.5$ (dashed orange line).}
    \label{fig:8}
\end{figure}

In Fig. \ref{fig:8} the coherence visibility is plotted as a function of the boson-boson interaction for $\varepsilon = 0$. 
At zero temperature, the visibility is maximal for vanishing interaction, when the ground state is the atomic coherent state. Increasing the absolute value of the interaction, the visibility decreases monotonically to zero. 
We observe that the loss of coherence occurs much more rapidly if the interaction is attractive. 
Indeed, a repulsive interaction tends to distribute the particles evenly on both sites.
As the interaction increases, tunneling weakens and the coherence slowly decreases because the state remains partially delocalized until the boson-boson interaction strongly dominates the tunneling energy, leading to a gradual freezing into a balanced configuration with $N/2$ particles on each site. On the contrary, even a modest attractive interaction is sufficient to tip the balance, favoring all the bosons to occupy one site and determining a rapid collapse of the coherence.

Finite temperature introduces non-trivial changes in the above picture. In the attractive regime, it produces an enhancement of the coherence visibility with respect to the zero-temperature value when $U$ is less than a certain (temperature-dependent) value \cite{Mazzarella_2012}. Furthermore, the maximal visibility no longer occurs for vanishing interaction. In the repulsive regime, the visibility is a non-monotonic function of the interaction strength, presenting a global maximum at a finite value of the interaction strength, which grows as the temperature increases. Quite interestingly, the same behavior is predicted to occur in three-dimensional Bose-Einstein condensates \cite{Vianello}. 

Introducing a small nonzero $\varepsilon$, the coherence visibility at $U=0$ is significantly reduced both at zero and finite temperature, while it remains almost unaffected for $|U|/JN \gg 0$. As a consequence, in the repulsive regime the visibility becomes a non-monotonic function of the interaction strength at all temperatures (including $T=0$), showing an initial increase before decreasing asymptotically to zero, whereas in the attractive regime the visibility remains a monotonically decreasing function of the modulus of the interaction strength.

\section{Summary and conclusion}\label{conclusion}

In this work we have studied the thermal equilibrium state of an atomic Josephson junction made of $N$ interacting bosons trapped by a quasi-one-dimensional asymmetric double-well potential. 
Since the interaction is short-ranged, the system can be successfully modeled by a two-site Bose-Hubbard Hamiltonian with an on-site energy bias $\varepsilon$. 
In the previous literature, the zero-temperature properties of the symmetric model ($\varepsilon = 0$) have been extensively explored, observing a crossover between a `Schrödinger cat' regime in the presence of large attractive interaction, a coherent superfluid-like regime in absence of interaction, and a Mott-like regime for large repulsive interaction. 
Furthermore, below a critical value of interaction in the attractive regime the ground state is quasi-degenerate, and it becomes exactly degenerate in the thermodynamic limit, where $U$ drives a quantum phase transition having the particle imbalance as order parameter.
Pitaevskii and Stringari have proposed a quantum phase model to describe the system in the thermodynamic limit, which allows to characterize the ground state through an effective one-body wavefunction obeying a Schrödinger equation with a suitable potential.

Here we have extended the analysis to the case of finite temperature and asymmetric wells, characterizing the mixed thermal state of the finite system by means of several complementary observables, such as the distribution of spectral weights, the thermodynamic and entanglement entropies, the population imbalance, the quantum Fisher information, and the coherence visibility, which is related to the fraction of condensed bosons. 

In summary, we have observed that finite temperature broadens the distribution of spectral weights, leading to larger thermodynamic and entanglement entropies, smaller population imbalance, smeared fluctuations of the relative number and relative phase, and decreased coherence visibility in the repulsive case. It is also at the origin of subtle effects such as a positive shift of the value of $U/J$ corresponding to maximum entanglement, the enhancement of the coherence visibility in the attractive regime, and the non-monotonic behavior of the visibility (and thus of the condensate fraction) in the repulsive regime. Such a detailed analysis of thermal effects is important to control decoherence, which negatively affects the performance of BJJs and similar systems in applications related to sensing and metrology \cite{Degen, Pezze_Rev, Ebgha}.

The on-site energy bias $\varepsilon$ explicitly breaks the left-right symmetry of the system and thus of the distribution of spectral weights, and its most evident effect is to induce a population imbalance in the wells. 
Other significant effects are the lifting of quasi-degeneracies, so that no quantum phase transition occurs in the thermodynamic limit, a positive shift of the value of $U/J$ corresponding to maximum entanglement, which can move from negative to positive, a strong reduction of the Fisher information in the attractive regime and of the coherence visibility at vanishing interaction. The inclusion of these effects brings the theory closer to a realistic implementation of a double-well potential, where the wells cannot be guaranteed to be perfectly symmetric.

In addition to the above exact numerical results, we have extended the quantum phase model to the asymmetric case; in Appendix \ref{sec:PS} we have discussed in detail how to compute its eigenvalues, eigenfunctions and thermal averages also at finite $N$. We have used this as a starting point to derive an effective semiclassical description for repulsive interaction and strong tunneling, i.e. in the interaction range $1/N \ll U/J \ll N$. This description provides analytical expressions for the thermal averages of observables having a classical analogue in terms of a modified Boltzmann weight with an effective `dressed' temperature, allowing valuable conceptual insights. For instance, we can characterize the junction capacitance in terms of the system parameters and explore its connection to the Fisher information via a fluctuation-dissipation relation. The approximate analytical results are typically in good agreement with the exact numerical calculations, especially when the observables do not depend on the relative phase of the two condensates or, in any case, when quantum fluctuations are suppressed due to a large number of particles ($N \gtrsim 100$). We have also discussed the possible limitations at very high temperature for certain observables.

The semiclassical description based on the effective temperature may find application in describing the thermal properties of other models for ultracold atomic gases confined in optical lattices. Within two-site models, there is the possibility of considering dipolar interactions \cite{Abad, Sowinski, Pizzardo} and, by increasing the number of semiclassical conjugate variables, multi-component systems with spin-orbit interactions \cite{Garcia}. For instance, our effective description could provide useful analytical guidance to distinguish the regime of coherence between wells and that of coherence between different species, in a case where obtaining a numerical solution is computationally expensive. A similar extension opens the possibility of studying the three-site Bose-Hubbard model \cite{Rab, Bradly, Olsen, Wilsmann} at finite temperature. In the integrable phase, where the model can be structured through two modes \cite{Wilsmann}, our description could be applied directly. We finally remark that on the technical side, possible extensions of this work could consider using a different type of coherent states \cite{Wimberger} or a different quantum phase model, see Appendix \ref{sec:PS}.

\section*{Acknowledgements}

CV and LS are supported by ``Iniziativa Specifica Quantum" of INFN and by the Project ``Frontiere Quantistiche" (Dipartimenti di Eccellenza) of the Italian Ministry of University and Research (MUR). 
MF acknowledges the SISSA and Massimo Capone for support.
LS is partially supported by funds of the European Union - Next Generation EU: European Quantum Flagship Project ``PASQuanS2", National Center for HPC, Big Data and Quantum Computing [Spoke 10: Quantum Computing]. LS also acknowledges the PRIN Project ``Quantum Atomic Mixtures: Droplets, Topological Structures, and Vortices" of MUR. 

\appendix
\section{Validity of the quantum phase model and possible improvements}\label{sec:PS}
The QPM is given by the Hamiltonian \eqref{HJ},
\begin{equation}\label{HJappend}
    \hat H_J = -U\frac{\partial^2}{\partial\theta^2} - i\varepsilon\frac{\partial}{\partial\theta} - JN \cos\theta,
\end{equation}
acting on the space of $2\pi$-periodic wavefunctions. In the case of symmetric wells ($\varepsilon =0$), the Schrödinger equation $\hat H_J \psi_n(\theta) = E_n \psi_n(\theta)$ takes the form $\psi_n''(\theta) + [E_n/U + (J N/U) \cos\theta]\psi_n(\theta) = 0$, which corresponds to the Mathieu equation ${\psi_n''(x) + [a-2q\cos(2x)]\psi_n(x) = 0}$ with $x=(\theta+\pi)/2$, $a = 4E_n/U$, and $q = 2JN/U$ \cite{Olver}. 
We are looking for solutions that are $2\pi$-periodic in $\theta$, hence $\pi$-periodic in $x$. 
The Mathieu equation admits two sets of $\pi$-periodic solutions, $\text{ce}_{2n}(x)$ (elliptic cosine) and $\text{se}_{2n}(x)$ (elliptic sine), labeled by even integers $2n = 0, 2, 4, \dots$, only if the parameter $a$ takes on the Mathieu characteristic numbers $a_{2n}(q)$ (for the elliptic cosine) and $b_{2n}(q)$ (for the elliptic sine). 
Therefore we can formally solve the Schrödinger equation as
\begin{equation}\label{mathieu}
    \psi_n(\theta) = \begin{cases}
        \frac{1}{\sqrt\pi}\,\text{ce}_{2n}\!\left(\frac{\theta+\pi}{2}\right), &  E_n = E_n^{(a)} = \frac{U}{4}\,a_{2n}\!\left(\frac{2JN}{U}\right),\\[2ex]
        \frac{1}{\sqrt\pi}\,\text{se}_{2n}\!\left(\frac{\theta+\pi}{2}\right), &  E_n = E_n^{(b)} = \frac{U}{4}\,b_{2n}\!\left(\frac{2JN}{U}\right).
    \end{cases}
\end{equation}
Both $a_{2n}(q)$ and $b_{2n}(q)$ are increasing functions of $n$, and $a_{2n}(q)\ge b_{2n}(q) \ge a_{2(n-1)}(q)$ for any $n, q >0$. 
Moreover $b_{2n}(q) \to a_{2n}(q)$ in the limits $q\to 0$ at fixed $n$ and $n\to \infty$ at fixed $q$. Notice that $b_0(q)$ is not defined, hence for $n=0$ we have only one solution, that is the ground state $\psi_0(\theta) = (1/\sqrt{\pi})\text{ce}_0(\frac{\theta+\pi}{2})$. 

The number of eigenstates $\psi_n(\theta)$ is in principle infinite, reflecting the fact that this approach is naturally suited to describe the system in the thermodynamic limit. Indeed, with canonical quantization the classical observable $k$, bounded in the interval $[-N/2, +N/2]$, is promoted to the unbounded operator $\hat k = -i\partial/\partial\theta$, and the mapping between the two can be made rigorous only in the limit $N\to\infty$. However, we can still apply the QPM to describe a system with a finite number $N$ of bosons, by introducing a cutoff on the label $n$ in such a way that the total number of eigenstates (i.e. the dimension of the Hilbert space) is equal to the dimension of the Hilbert space of \eqref{HBH}, which is $N+1$. If $N$ is even, taking into account the ordering of the eigenvalues in terms of Mathieu characteristic numbers, this means that $n$ must take on the integer values from $0$ to $N/2$. Thermal averages may then be computed as
\begin{equation}\label{PSave}
    \langle \hat Q \rangle = \frac{1}{Z}\sum_{n=0}^{\frac{N}{2}}e^{-\beta E_n^{(a)}}Q_n^{(a)} +\frac{1}{Z}\sum_{n=1}^{\frac{N}{2}}e^{-\beta E_n^{(b)}}Q_n^{(b)},
\end{equation}
where 
\begin{subequations}
\begin{align}
    Q_n^{(a)} &= \int_{-\pi}^{\pi}\frac{d\theta}{\pi}\,{\textstyle\text{ce}_{2n}\!\left(\frac{\theta+\pi}{2}\right)}\braop{\theta}{\hat Q}{\theta}\,{\textstyle\text{ce}_{2n}\!\left(\frac{\theta+\pi}{2}\right)},\\
    Q_n^{(b)} &= \int_{-\pi}^{\pi}\frac{d\theta}{\pi}\,{\textstyle\text{se}_{2n}\!\left(\frac{\theta+\pi}{2}\right)}\braop{\theta}{\hat Q}{\theta}\,{\textstyle\text{se}_{2n}\!\left(\frac{\theta+\pi}{2}\right)},
\end{align}
\end{subequations}
and $Z = \sum_{n=0}^{N/2} e^{-\beta E_n^{(a)}} + \sum_{n=1}^{N/2} e^{-\beta E_n^{(b)}}$. If instead $N$ is odd, the index $n$ would take on the integer values between $0$ and $\lfloor N/2 \rfloor$ in the $a$-summations, and the integer values between $1$ and $\lceil N/2 \rceil$ in the $b$-summations.

For $\varepsilon\neq 0$, the Schrödinger equation $\psi_n''(\theta) + (i\varepsilon/U)\psi_n'(\theta) +[E_n/U + (J N/U) \cos\theta]\psi_n(\theta) = 0$ does not have an exact solution in terms of known functions. 
Since we are looking for $2\pi$-periodic solutions, we may expand $\psi_n(\theta)$ in the Fourier basis $\{e^{ij\theta}\}_{j\in\mathbb Z}$, taking into account that the dimensionality of the Hilbert space is restricted to $N+1$. 
Thus, for $N$ even, $\psi_n(\theta) = (1/\sqrt{2\pi})\sum_{j=-N/2}^{N/2} f^{(n)}_j e^{ij\theta}$, with $\Vert \vec f^{(n)}\Vert^2 = 1$, and the Schrödinger equation becomes $\mathbf M \vec f^{(n)} = (4E_n/U)\vec f^{(n)}$, where $\mathbf M$ is a $(N+1)\times(N+1)$ tridiagonal matrix with elements $M_{jj} = 4\varepsilon j/U+4j^2$, $M_{j(j\pm1)} = -2JN/U$. Thermal averages may then be computed as 
\begin{align}\label{meas}
    \langle \hat Q\rangle =&\, \frac{1}{Z}\sum_{n=0}^N e^{-\beta E_n}\times\nonumber\\
    &\sum_{j,\ell=-\frac{N}{2}}^{\frac{N}{2}}f_j^{(n)*}f_\ell^{(n)}\int_{-\pi}^{\pi} \frac{d\theta}{2\pi}\,e^{-ij\theta}\braop{\theta}{\hat Q}{\theta}e^{i\ell\theta}.
\end{align}

In light of the above discussion and the results presented in the previous sections, we can now make some remarks on the range of validity of the QPM. The naive range of validity suggested by the corresponding classical Hamiltonian \eqref{Hcl} is $|U|/J \gg 1/N$.  While this condition is sufficient to obtain quantitatively accurate results for quantities that depend on the relative number $\hat k$ (e.g. population imbalance and Fisher information), it is not sufficient to accurately compute quantities that depend on the relative phase $\hat\theta$. For instance, when the number of particles is small ($N\lesssim 100$), the QPM systematically underestimates the coherence visibility even if $|U|/J \gg 1/N$. For $N\sim100$, accordance with the exact result is satisfactory, as the relative error is less than 1\%. 
This is consistent with what we find for the energy gap between the first excited state and the ground state $E_1-E_0$, which in the QPM is given by $\hbar\omega_J\sqrt{\langle \cos^2\hat\theta\rangle/\langle \cos\hat\theta\rangle}$ \cite{Pitaevski_Stringari}; this expression systematically underestimates the exact result, and the accordance becomes satisfactory only for $N \gtrsim 100$. This reflects the fact that the QPM is intrinsically consistent with the thermodynamic limit, where it matches with Eq. \eqref{HBH}, while introducing the cutoff based on the dimension of the Hilbert space of \eqref{HBH} may give difficulties for observables depending on the relative phase if the system is made of few bosons.

The shortcomings of the QPM at small $N$ can be addressed by the improved quantum phase model (IQPM) derived by Anglin \emph{et al.} \cite{Anglin_QPM} using the Bargmann representation, which can be extended to asymmetric wells as
\begin{align}\label{IQPM}
    \hat H_\text{IQPM} = &-U\frac{\partial^2}{\partial\theta^2} - i\varepsilon\frac{\partial}{\partial\theta} - JN\left(1+\frac{1}{N}\right)\cos\theta\nonumber\\
    &- \frac{J^2}{2U}\cos 2\theta.
\end{align}
The construction of Anglin \emph{et al.} also shows that by defining appropriately the scalar product on the IQPM wavefunctions, one projects out the unphysically high Fourier components in the wavefunctions, limiting the dimension of the physical Hilbert space to $N+1$. 

The IQPM could then be used as the basis for the semiclassical approximation; thermal averages would be computed with the statistical weight $e^{-H_\text{IQPM}/k_B T_\text{eff}}$, where $H_\text{IQPM} = Uk^2 + \varepsilon k -JN(1+\frac{1}{N})\cos\theta-(J^2/2U)\cos 2\theta$ and $T_\text{eff} = (\hbar \omega_\text{IQPM}/2k_B)\coth(\hbar\omega_\text{IQPM}/2k_B T)$, with the Josephson frequency
\begin{equation}\label{omega_i}
    \omega_\text{IQPM} = \frac{\sqrt{2UJN}}{\hbar}\sqrt{1 + \frac{1}{N}+\frac{2J}{UN}}.
\end{equation}
This however provides only a slight improvement with respect to the semiclassical approximation based on the frequency $\omega_J = \sqrt{2UJN}/\hbar$ [Eq. \eqref{omegaj}]. 

In summary, while at the quantum level the IQPM \eqref{IQPM} provides an improved effective description of the two-site Bose-Hubbard model \eqref{HBH} with respect to the QPM \eqref{HJappend}, the resulting `bare' semiclassical approximations for thermal averages are not significantly different. The QPM-based semiclassical approximation we presented in the main text of the paper, since it includes first-order quantum corrections in the Josephson frequency and thus in the effective temperature, is clearly more accurate than the bare IQPM-based semiclassical approximation. Significant progress could be achieved by including similar quantum corrections in the frequency \eqref{omega_i}, e.g. by applying the methods of Ref. \cite{Furutani} to the IQPM \eqref{IQPM}, obtaining a modified version of Eq. \eqref{Omegaj}. This would represent an interesting technical extension of the present work.

\bibliography{References}

\end{document}